\documentclass[12pt,preprint]{aastex}

\shorttitle{HII Region Turbulence}
\shortauthors{Spangler et al}

\begin{document}

\title{Analytic Estimates of the Effect of Plasma Density Fluctuations on HII Region Density Diagnostics}
\author{Steven R. Spangler and Brandon M. Bergerud}
\affil{Dept. of Physics and Astronomy, University of Iowa}
\author{Kara M. Beauchamp}
\affil{Dept. of Physics and Engineering, Cornell College}
\begin{abstract}
An analytic calculation is made of the effect of plasma density fluctuations on some spectroscopic diagnostics commonly used in the study of HII regions and planetary nebulae. To permit an analytic treatment, attention is restricted to the case of density fluctuations possessing an exponential probability distribution function (pdf).  The present investigation is made in support of a completely numerical and more extensive study of nebular diagnostics by \cite{Bergerud19}. Results from this paper are presented in terms of graphs of the observed quantity (spectroscopic line ratio) versus mean nebular density.  Our results yield a higher density estimate, given the same observed line ratio, for the case of a nebula with density fluctuations than for the case of a nebula with uniform density. This is qualitatively consistent with the typically observed case, in which the observations lead to the inference of a filling factor $< 1$.  Our results are in quantitative agreement with those of \cite{Bergerud19}, and thus corroborate those calculations for the case of an exponential pdf.  
\end{abstract}

\section{Introduction}
This paper is a supplement to that of \cite{Bergerud19}.  To allow it to exist as an independent document, we briefly summarize the motivation and goals of \cite{Bergerud19}.

\cite{Bergerud19} consider the effect of plasma turbulence on the classical spectroscopic diagnostics used for HII regions and planetary nebulae (\cite{Osterbrock89,Draine11}).  Turbulent fluctuations in density, magnetic field, plasma flow velocity, etc., are known to be present in plasmas such as the solar corona, solar wind, and the Warm Ionized Medium (WIM) component of the Interstellar Medium (ISM).  This turbulence may be reasonably assumed to be present in HII regions and planetary nebulae as well.

The goal of \cite{Bergerud19} is to investigate the consequences of such turbulence for spectroscopically-inferred values of mean density, mean temperature, and Abundance Discrepancy Factor (ADF) in these nebulae.  One of the main goals of \cite{Bergerud19} is to determine if turbulent density fluctuations, characterized by a specified probability density function (pdf), can quantitatively account for ``filling factors'' substantially less than unity.  An assumption of the analysis of \cite{Bergerud19} is that turbulent density fluctuations with a plausible pdf are a more natural model to have of a nebula than the classical view of a filling factor in which the nebula has clouds of uniform density suspended in a vacuum.

The study of \cite{Bergerud19} is numerical, in that simulated nebulae are created in a computer and produce simulated observables.  The purpose of this paper is to serve as a check on some of those numerical results, and also to gain possible physical insight from an analytic treatment.  To allow an analytic approach, we restrict ourselves to a discussion of the filling factor in the case in which the only relevant turbulent fluctuations are those of density, and further restrict ourselves to density fluctuations possessing an exponential pdf.  

\section{Approximate Atomic Energy Level Model}
Atomic energy levels that are useful as density diagnostics are those from ions with a $np^3$ electron configuation (Draine 2011).  Several such ions and their transitions are given in Table 1 of Bergerud, Spangler, and Beauchamp (2019).  The analysis presented here envisions a 3 level atom, consisting of a ground state (0), and 2 nearly degenerate excited states (1 and 2, to be identified with the $^2D$ states).  The Einstein A values for the transitions are not the same, $A_{10} \neq A_{20}$ We make the following simplifying assumptions. 
\begin{enumerate}
\item Only transitions between the fine structure states 1 and 2 and the ground state occur, i.e $A_{10} \neq 0, A_{20} \neq 0, A_{21} = 0$.
\item The excited states are only weakly excited, i.e. $n_1, n_2 \ll n_0$, where $n_1$ and  $n_2$ are number densities of ions in the 1 and 2 states. 
\end{enumerate} 

In this case, the emission coefficients in the two transitions $1 \rightarrow 0 \equiv 1$ and  $2 \rightarrow 0 \equiv 2$ are
\begin{eqnarray}
\epsilon_1 = \left( \frac{h \nu_{1}}{4 \pi} \right) n_1 A_{10} \\
\epsilon_2 = \left( \frac{h \nu_{2}}{4 \pi} \right) n_2 A_{20}
\end{eqnarray}
The limitations of this model are that higher excited states (specifically the $^2P$ states) are ignored.  Furthermore, radiative and collisional transitions between the 1 and 2 states are ignored.  

Applying the usual assumption of detailed balance to the equilibrium level populations gives the following expression for population (number of atoms unit volume) in the 1 state, and a similar expression for the 2 state,
\begin{equation}
n_1 = \left[ \frac{n_e q_{01}}{A_{10} (1 + \frac{n_e q_{10}}{A_{10}})} \right] n_0
\end{equation}
where  $ n_1$ and $ n_0$ are the number densities of the 1 state and the ground state, respectively, $ q_{01}$ is the collision frequency between states 0 and 1 and  $q_{10}$ is the collision frequency for the downward collisional transition.  $A_{10}$ is the Einstein coefficient for the radiative transition.  Equation (3) can be used in Equation (1) for the appropriate emission coefficient, and the same done for the transition $2 \rightarrow 0$.

Equations (1) and (2) give the expressions for the spectral line emission coefficient, which is a function of the electron density $n_e$ and the ground state density $n_0$.  In a medium with density fluctuations, both of these will vary with position, as will $n_1$ and $\epsilon_1$.  We further simplify matters by assuming that the nebula is nearly pure hydrogen.  In this case, using assumption (2) above, $n_0 = X n_p = X n_e$ where $X$ is the abundance of the ion relative to hydrogen and $n_p$ is the number density of protons. Determination of $X$ would require not only knowledge of the elemental abundance, but of the ionization state as well.

\section{Density Probability Distributions and the Exponential PDF}
The purpose of this investigation is to study the effect on various radiative plasma diagnostics when the plasma density (as well as other plasma parameters) varies in a stochastic manner through the nebula.  The density is described by a probability density function (pdf) $p(n)$.  This is obviously defined such that $p(n)dn$ is the incremental probability of the density being in the range $n \rightarrow n +dn$.  Our approach is to first derive general expressions for quantities such as the emission measure or spectral line intensity that are valid for any pdf, and then to specialize to the case of an exponential pdf.  The exponential distribution is chosen for reasons of analytic convenience.

Properly normalized, the exponential pdf is described by
\begin{equation}
p(n) = a e^{-n/n_0} = a e^{-an}\mbox{ , } a \equiv \frac{1}{n_0}
\end{equation}
With this expression,  $<n> = n_0$.

\section{Expressions for Radiative Diagnostics}
The measured quantities of interest for determining the filling factor of an HII region or planetary nebula are the radio brightness temperature due to thermal bremstrahlung emission and the intensity of spectral line emission due to transitions from a density-sensitive excited state to a ground state.  In either case, the intensity in the optically-thin case is given by
\begin{equation}
I(\nu) = \int_0^L dz \epsilon(\nu,z)
\end{equation}
 
where $\epsilon$ is the emission coefficient, z is the coordinate along the line of sight and $\nu$ is the frequency of the emission.  In the simulations of \cite{Bergerud19}, the HII region is approximated by a number $N$ of independent cells along the line of sight, so Eq (5) can be represented as 
\begin{equation}
I(\nu) = \sum^N_{i=1} \Delta z \epsilon_i(\nu) = \Delta z  \sum^N_{i=1}  \epsilon_i(\nu) = N \Delta z \left[ \frac{ \sum^N_{i=1}  \epsilon_i(\nu)}{N} \right] 
\end{equation}
with $N$ being the number of cells along the line of sight. If we want the mean value of $I$, we take the expectation value and get 
\begin{equation}
<I> = L <\epsilon(n)>
\end{equation}
Observationally, $<I>$ can be thought of as the average of several apparently equivalent lines of sight through an HII region. Equation (7) is the basic expression we use; it relates the mean value of the intensity to the expectation value of $\epsilon$.  In our study of model nebulae with only density fluctuations, we assume $\epsilon$ is varying because the plasma density varies within the HII region, so 
\begin{equation}
<\epsilon(n)> = \int^{\infty}_0 dn p(n) \epsilon(\nu,n)
\end{equation}
 In the discussion here, we do not consider the effect of temperature fluctuations on $\epsilon$ (see Section 2.3 of Bergerud et al 2019). Equation (8) can be evaluated (analytically or numerically) for any pdf, and several plausible ones are considered in \cite{Bergerud19}.  
\subsection{Statistics of Radio Continuum Emission} 
The emission coefficient for thermal Bremsstrahlung is taken from Rybicki and Lightman (1979), and can be parameterized as 
\begin{equation}
\epsilon_{ff} = \frac{A}{\sqrt{T}}n_e^2
\end{equation}
where $A$ contains fundamental physical constants, the velocity-averaged Gaunt factor, the frequency, etc. As stated above, in this analysis we assume that the temperature is a constant. The result is that 
\begin{equation}
<I> = \frac{AL}{\sqrt{T}} <n^2> =  \frac{AL}{\sqrt{T}} \int^{\infty}_0 dn p(n) n^2
\end{equation}

\subsubsection{Mean Radio Brightness Distribution for Exponential Statistics }
For the case of an exponential pdf, 
\begin{equation}
<n^2> =  \int^{\infty}_0 dn p(n) n^2 = a \int^{\infty}_0 dx e^{-ax} x^2 = a \left(\frac{2!}{a^3} \right) = 2 n_0^2  
\end{equation}
This expression for $<n^2>$ can be substituted into Eq (6).  The obvious conclusion is that the density estimate obtained by a radio continuum measurement is $\sqrt{<n^2>} = \sqrt{2} n_0$.  In the case of a nebula with random density fluctuations,  the measured, inferred density is higher than the true mean value.  
\subsection{Statistics of the Ratio of Spectral Lines}
We now consider the density inferred from the ratio of spectral line intensities, and how that compares with the value from a radio continuum measurement. We use the model for an ionic energy level diagram described in Section 1 above, and Equations (1) - (3). 

Since the atomic physics parameters occur in multiplicative combinations, we use some shorthand notation.  Let 
\begin{eqnarray}
D_1 \equiv \left( \frac{h \nu_{1}}{4 \pi} \right) A_{10} q_{01} (T)  \\
c_1 \equiv \frac{q_{10}(T)}{A_{10}}
\end{eqnarray}
with corresponding expressions for transitions from excited state 2.  
Equations (12) and (13) express the fact that $D_1 \mbox{ and } c_1$ are functions of the temperature T. Then 
\begin{eqnarray}
\epsilon_1 = D_1 X \frac{n_e^2}{(1 + c_1 n_e)} \mbox{ and }  \\
\epsilon_2 = D_2 X \frac{n_e^2}{(1 + c_2 n_e)}
\end{eqnarray}
where $X$ is defined in Section 1. 
  
With those expressions, we proceed to the expression for the mean value of the emission coefficient, and expectation value of the intensity in each line. 
\begin{eqnarray}
\left< \epsilon_1 \right> = D_1 X \left< \frac{n_e^2}{1 + c_1 n_e} \right> =  D_1 X \int^{\infty}_0 dn p(n) \left( \frac{n^2}{1 + c_1 n} \right)  \mbox{ so }  \\
\left< I_1 \right> = (D_1 X L) \int^{\infty}_0 dn p(n) \left( \frac{n^2}{1 + c_1 n} \right)  \mbox{ and, similarly }  \\
\left< I_2 \right> = (D_2 X L) \int^{\infty}_0 dn p(n) \left( \frac{n^2}{1 + c_2 n} \right) 
\end{eqnarray}
where we have also employed the assumption of a hydrogen nebula for which $n_e = n_p = n$.
These expressions, and their relation to the  $\sqrt{<n^2>}$ quantity measured in radio continuum measurements, depends on the statistics of the density fluctuations. The statistical quantity to be evaluated is 
\begin{equation}
\left< \frac{x^2}{1 + cx} \right> = \int^{\infty}_0 dx p(x) \left( \frac{x^2}{1 + c x} \right) 
\end{equation}
for the pdf of choice. 
\subsubsection{Density-Sensitive Emission Line Ratios for Exponential Density Fluctuations}

We substitute Eq (4) into (19).  We also change variables of integration from x (or n) to $y \equiv cx$.  This change of variables then produces the following expression for the line intensity $<I  >$, dropping the subscript for the time being; the same equation applies to $<I_1>$ and $<I_2>$.  
\begin{eqnarray}
<I> = \frac{DXL}{c^2} \left[ g \int_0^\infty dy e^{-gy} \left( \frac{y^2}{1+y} \right) \right] \\
g \equiv \frac{1}{c n_0}
\end{eqnarray}
In Equations (20) and (21), $c$ is the atomic physics coefficient defined in Equation (13), not the speed of light. We define the term in square brackets as a function $H(g)$, i.e.
\begin{equation} 
H(g) \equiv  g \int_0^\infty dy e^{-gy} \left( \frac{y^2}{1+y} \right)
\end{equation}

There is an analytic expression for $H(g)$, 
\begin{equation} 
H(g) \equiv \frac{1-g}{g} +g e^g \Gamma[0,g]
\end{equation}
where $\Gamma[a,b]$ is the incomplete Gamma function.  However, for the analysis carried out here we use a  {\em Mathematica} notebook to define $H(g)$ as a function. A plot of $H(g)$ is shown in Figure 1.  

\begin{figure}[h]
\epsscale{0.40}
\includegraphics[width=20pc]{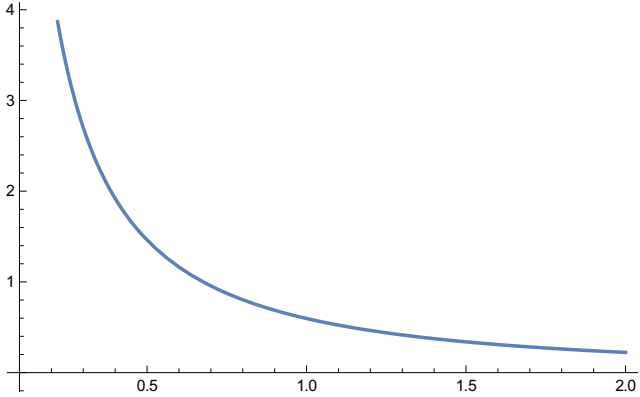}
\vskip-.15truein
\caption{Plot of the function $H(g)$, that describes the intensity of a collisionally-excited line in a 3-state atom or ion. Abscissa is $g$ and the ordinate is $H(g)$.}
\end{figure}
This plot shows the expected result that as $g$ decreases ($n_0$ increases), the mean intensity of the line increases.  
\subsubsection{Expression for Line Ratios, Exponential PDF}
The diagnostic measurement we use is the ratio of 2 lines, each due to a transition from an excited state to the ground state.  So we are interested in the quantity $\frac{<I_1>}{<I_2>}$.  Both of the intensities are described by expressions of the form in Eq (20), 
\begin{equation}
<I> = \frac{DXL}{c^2} H(g)
\end{equation}
For the two transitions, D, c, and therefore g, will be different, i.e. we have $D_1$ and  $D_2$, $c_1$ and  $c_2$,  $g_1$ and  $g_2$, although the underlying density pdf (Eq (4)) is the same.  

The intensity ratio is then given by
\begin{equation}
\frac{<I_1>}{<I_2>} = \frac{D_1 X L}{c_1^2} \frac{c_2^2}{D_2 X L} \frac{H(g_1)}{H(g_2)} = \frac{D_1}{D_2} \left( \frac{c_2}{c_1} \right)^2 \frac{H(g_1)}{H(g_2)}
\end{equation} 
where $g_1 = \frac{1}{c_1 n_0}$ and $g_2 = \frac{1}{c_2 n_0}$.  \\

The values of $g_1$ and  $g_2$ are related (see above expression).  
\begin{equation}
g_2 = \frac{1}{c_1 \left( \frac{c_2}{c_1} \right) n_0} = \left( \frac{c_1}{c_2} \right) g_1 = {\cal X} g_1 
\end{equation}
where ${\cal X} \equiv \left( \frac{c_1}{c_2} \right)$. ${\cal X}$ can be either $\leq 1$ or $> 1$.  
With all of this, we can obtain an expression for the line ratio $\frac{<I_1>}{<I_2>}$ as a function of one independent variable, $g_1$.  
\begin{equation}
\frac{<I_1>}{<I_2>} = \left( \frac{D_1}{D_2} \right) \frac{1}{{\cal X}^2} \frac{H(g_1)}{H({\cal X}g_1)}
\end{equation}

\subsection{Line Ratios in the Case of Osterbrock-Flather Statistics}
We now carry out the same calculation for the Osterbrock and Flather (1959) model for the density statistics, which consists of a dual delta function pdf.  
Begin with a slight variation of Eq (20) in which we change variables from $n \rightarrow y \equiv nc$
\begin{equation}
<I_1> = \frac{D_1XL}{c_1^3} \int_0^{\infty} dy p(n(y) = \frac{y}{c_1}) \left( \frac{y^2}{1 + y}  \right)
\end{equation}
For the Osterbrock-Flather model, 
\begin{equation}
p(n) = (1-f) \delta(n) + f \delta(n - n_d)
\end{equation}
where $n_d$ is the density in the droplets.  This gives
\begin{equation}
<I> = \frac{DXL}{c^3} f \int_0^{\infty} dy \delta(\frac{y}{c} - n_d)\left( \frac{y^2}{1+y} \right)
\end{equation}

so 
\begin{eqnarray}
<I> = \frac{DXL}{c^2} f \int_0^{\infty} dy \delta(y - cn_d)\left( \frac{y^2}{1+y} \right) \\
<I> = \frac{DXL}{c^2} f \left( \frac{c^2 n_d^2}{1 + cn_d} \right)
\end{eqnarray}
Switching back to the case of 2 spectral lines, we can use Equation (32) to obtain an expression for the ratio of line intensities in the same notation as that for the case of exponentially-distributed density fluctuations.  
\begin{equation}
\frac{<I_1>}{<I_2>} = \frac{D_1}{D_2}  \left[ \frac{(1 + \frac{1}{ {\cal X} g_1})}{(1 + \frac{1}{g_1})} \right]
\end{equation}
Equation (33) is the Osterbrock-Flather counterpart to Equation (27) for the case of an exponential pdf of density fluctuations. One distinction to be noted is that in the case of the exponential pdf, $g_1 \equiv \frac{1}{c_1 n_0}$, where $n_0$ is the mean density.  In the case of Osterbrock-Flather statistics, $g_1 \equiv \frac{1}{c_1 n_d}$, where $n_d$ is the density in the clouds embedded in a vacuum.

It should be emphasized that Equation (33) is also valid for the case of a nebula with uniform density, since in the Osterbrock-Flather model the component with zero density contributes no light.  When line ratios are taken, the filling factor cancels.  

\section{Choice of Atomic Physics Parameters}
Our expressions for the line intensity ratios (the spectroscopic density diagnostic), Equation (27) (for exponential pdf) or Equation (33) (Osterbrock-Flather model) depend on atomic physics parameters such as $A_{10}$, $q_{02}$, etc.  These determine the  parameters $D_1/D_2$ and $\cal X$, as well as the mapping from plasma density to the variable $g_1$. Values for these atomic physics parameters, and corresponding coefficients in Equation (27) and (33) may be selected for real line ratios, such as SII[6716/6731].  When this is done with the real atomic data (e.g. from Osterbrock 1989 or Draine 2011), the corresponding curves for $\frac{<I_1>}{<I_2>}$ qualitatively resemble the plots in Osterbrock (1989) and Draine (2011), but show poor agreement in detail.  That is, the line ratios in the asymptotic high and low density limits have the wrong values, and the density at which the ``crossover'' occurs is not correct.  This indicates that the approximations enumerated in Section 1 are not an accurate representation of the true situation, involving 5 states (one ground state and four excited states), with collisional and radiative transitions permitted between all states.  This means that the formulas derived above, and used subsequently, constitute an approximation scheme for estimating the effect of turbulence on observed line ratios.

Given this situation, our approach was to use Equations (27) and (33), but to adjust the parameters $D_R \equiv \frac{D_1}{D_2}$, $\cal X$, and $c_1$ to produce a curve that closely resembled the function plotted in Figure 5.3 of Osterbrock (1989) for the line ratios of SII(6716/6731).  That is, with adjusted values of these three parameters, the line ratio  $\frac{<I_1>}{<I_2>}$ had the proper functional dependence on $n$ that had the correct asymptotic limits at low and high densities, and made the transition between asymptotic limits at approximately the right density.  This permits us to use Equation (27) to explore the effect of density turbulence on the observed line ratio.

Table 1 illustrates the magnitude of the corrections necessary to simulate the true density dependence of the SII(6716/6731) line ratio, given the approximations explicit in Equations (27) and (33). The upper row gives values calculated from the correct atomic physics parameters given in Osterbrock (1989) and Draine (2011).  The lower row gives adjusted parameters that reproduce the behavior of the SII(6716/6731) ratio plotted in Figure 5.3 of Osterbrock (1989).\\

\begin{tabular}{||l|l|l|l||} \hline
 Source & $D_R$ & $\cal X$ & $c_1$ (cm$^3$)\\
\hline 
  Osterbrock (1989) & 0.295 & 3.38 & $2.37 \times 10^{-4}$ \\
  Adjusted  & 1.42 & 3.38 & $1.11 \times 10^{-3}$ \\
  \hline
\end{tabular}

\section{Analytic Results for Filling Factor}

We can now combine the results above to obtain equations giving the inferred filling factor in the case of an exponential pdf.  We plot together the expressons for $\frac{<I_1>}{<I_2>}$ given in Equations (27) and (33), and using the ``adjusted'' atomic physics parameters in Table 1.  The results are shown Figure 2.   
\begin{figure}[h]
\epsscale{0.40}
\includegraphics[width=20pc]{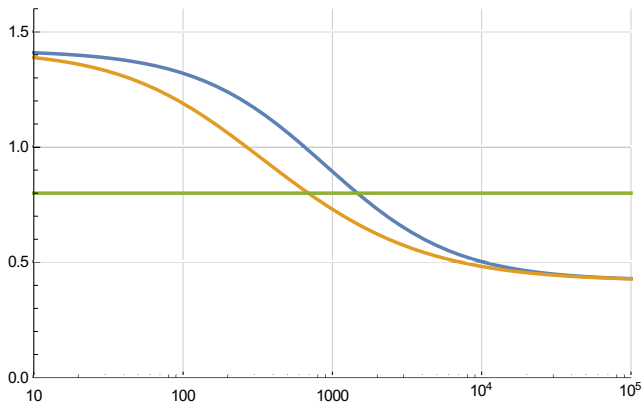}
\vskip-.15truein
\caption{Line ratio $\frac{<I_1>}{<I_2>}$ for SII(6716/6731) as a function of nebular density for case of Osterbrock-Flather statistics (blue curve) and exponential density fluctuations (orange curve). The abscissa is nebular density (cm$^{-3}$) and the ordinate is observed line ratio. In the case of the Osterbrock-Flather picture, the density is the density within the plasma-containing clouds.   The horizontal line indicates a fixed observed value for the line ratio.   }
\end{figure}

Once again, the Osterbrock-Flather expression is used to show the observed line ratio for the case of a model uniform nebula as well as one with uniform clouds immersed in a vacuum.  This diagram illustrates that the same, observed line ratio (horizontal green line) is obtained for a uniform nebula, or nebula with a small filling factor possessing ``clouds'' of density $n_d$, or by a nebula with an exponential density pdf and a mean density $ n_0 < n_d$.  Figure 2 also shows that the degree to which these two density values disagree depends on $n_0$, being smaller for larger densities.

Obtaining a compact analytic expression for $n_0(n_d)$ would require equating Equations (27) and (33) and then inverting the function $H(g)$.  In lieu of this, we use a graphical scheme based on Figure 2.

\begin{figure}[h]
\epsscale{0.40}
\includegraphics[width=20pc]{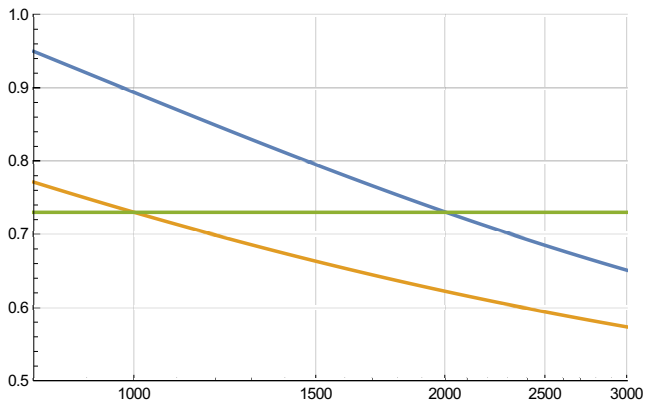}
\vskip-.15truein
\caption{Plot same as Figure 2, except for a narrower range of nebular densities and observed line intensity ratios. The horizontal line indicates a fixed observed value for the line ratio of $\frac{<I_1>}{<I_2>} = 0.73$.   }
\end{figure}

Figure 3 shows an expanded version of Figure 2, illustrating the conclusions of our analysis for an arbitrarily-chosen value of the observed line ratio $\frac{<I_1>}{<I_2>} = 0.73$.

Figure 3 shows that, in the case of an observed line ratio of $\frac{<I_1>}{<I_2>} = 0.73$, an analysis assuming a uniform nebula, or an Osterbrock-Flather model would yield a plasma density of 2000 cm$^{-3}$. This density would pertain either to the entire nebula for the uniform nebula assumption, or the density within the ``clouds'' in the case of the Osterbrock-Flather picture.  However, Figure 3 shows that the same line ratio would occur in the case of nebula with a mean density of 1000 cm$^{-3}$ and fluctuations in density described by an exponential pdf.

This discrepancy becomes more pronounced when the mean density is lower, as illustrated in Figure 4.
\begin{figure}[h]
\epsscale{0.40}
\includegraphics[width=20pc]{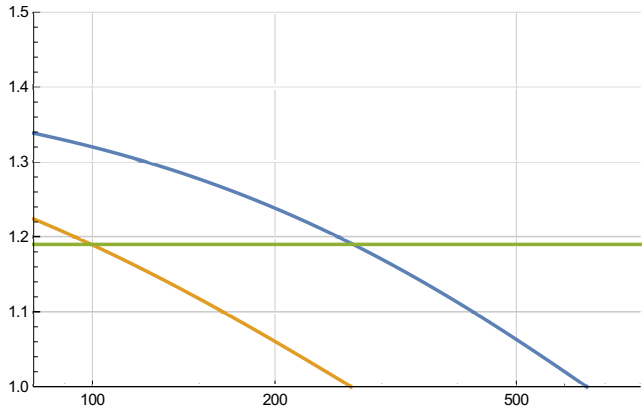}
\vskip-.15truein
\caption{Plot same as Figures 2 and 3, except for a narrower range of nebular densities and observed line intensity ratios. The horizontal line indicates a fixed observed value for the line ratio of $\frac{<I_1>}{<I_2>} = 1.19$.   }
\end{figure}

Figure 4 shows results in the case of an observed line intensity ratio of $\frac{<I_1>}{<I_2>} = 1.19$.  In this case, the uniform nebula or Osterbrock-Flather model would give a density of 270 cm$^{-3}$.  However, the same observed line ratio could be reproduced for a nebula with exponentially-distributed density fluctuations, and a mean density of 100 cm$^{-3}$.  Thus the difference between the two estimates is greater in the case of lower densities, in agreement with a conclusion from \cite{Bergerud19}.
\subsection{Quantitative Values for Filling Factors}
The results presented immediately above show that the same observable line intensity ratio may be produced either by a nebula with an exponentially distributed set of density fluctuations, or a nebula with a uniform plasma density in the emitting regions, and a substantially higher density.  Qualitatively, this is not surprising, although our results place this on a quantitative basis.

The observational evidence for a nebular filling factor substantially less than unity comes from a comparison of two density estimates, one from measurement of spectroscopic line ratios, and the other from an emission measure measurement, such as provided by radio continuum measurements.  It is obvious in the case of the Osterbrock-Flather model that the latter will be smaller than the former, and reported values of filling factors in the literature have used the filling factor to reconcile the two measurements.  We now wish to consider what values of filling factor would occur in the case of the exponential pdf, and how they compare with published estimates. We utilize results presented in the previous section.

Equation (11) shows that for the case of an exponential pdf, the root-mean-square density obtained from a radio continuum brightness temperature will exceed the mean value $n_0$ by a factor of $\sqrt 2$.  For the model nebula in Figure 3,
$n_0 = 1000 \mbox{ cm}^{-3}$, so a radio continuum measurement would estimate an rms density of $1410 \mbox{ cm}^{-3}$.  Use of the SII(6716/6731) line ratio would yield a density of 2000 cm$^{-3}$, so in this case
\begin{eqnarray}
  \frac{n_e(EM)}{n_e(CEL)} = \frac{1}{\sqrt{2}}  \\
  f = (\frac{n_e(EM)}{n_e(CEL)})^2 = 0.50
\end{eqnarray}

This is exactly the result obtained by the simulations presented in Section 3.1 and illustrated in Figure 5 of Bergerud, Spangler, and Beauchamp (2019), and thus serves as an independent check of the results presented there.

For the lower density nebula for which a line ratio of 1.19 is measured, the corresponding values are $n_e(EM) = 141$  cm$^{-3}$ and $n_e(CEL) = 270$  cm$^{-3}$, so
\begin{eqnarray}
  \frac{n_e(EM)}{n_e(CEL)} = \frac{141}{270}  \\
  f = (\frac{n_e(EM)}{n_e(CEL)})^2 = 0.27
\end{eqnarray}

These filling factors, obtained from our analytic approach, span the range of those discussed in the context of the Rosette Nebula by Costa et al (2016).  This indicates that these results are immediately applicable to nebulas in which the filling factor is less than unity by slight to moderate degrees.

However, these calculations also give qualitative support for the simulation results of Bergerud, Spangler, and Beauchamp (2019) (e.g. Figure 6 of  Bergerud, Spangler, and Beauchamp (2019)), which show that much smaller filling factors, that explain a larger portion of published results, occur in the case of density pdfs with more pronounced tails than the exponential function, such as lognormal or Pareto distribution pdfs.

\end{document}